\documentclass[12pt,letterpaper]{article}
\usepackage{graphicx} 
\usepackage{cite}

\begin{document}

\raggedbottom 

\pagestyle{plain}

\title{Various approaches to solving nonlinear equations}

\author{John C. Nash and Ravi Varadhan}

\date{2024-05-31}

\maketitle

\abstract{%
Modelling real world systems frequently requires the solution of systems of nonlinear equations. A number of approaches have been suggested and developed for this computational problem. However, it is also possible to attempt solutions using more general nonlinear least squares or function minimization techniques. There are concerns, nonetheless, that we may fail to find solutions, or that the process will be inefficient. Examples are presented with R with the goal of providing guidance on the solution of nonlinear equations problems.
}

\section{Background}\label{background}

Mathematical modelling translates real-world phenomena into equations and
inequalities, the solution of which allows us to explain and predict the
behaviour of the real-world system of interest. This is illustrated by
an example of from Nelson and Hodgkin (\cite{NelsonHodgkin81}{1981}) that has been translated to a test
problem for nonlinear equations
and nonlinear least squares in John E. Dennis, Gay, and Vu (\cite{DenGayVu1983}{1983}).

We will write our nonlinear equations in the form

\[  r(x)  =  0 \]
where \(x\) are the \textbf{parameters} we wish to determine by solving the
equations that the \textbf{residuals} \(r(x)\) provide. Usually we anticipate there will
be an equal number \(n\) of parameters and equations. However, there can be
systems with no solution, or sets of \(m>n\) equations which satisfy the
equations. Alternatively, minimizing the \textbf{sum of squares} \({r(x)}^2\)
provides an approach to solving the nonlinear equations if we can find a
solution with the sum of squares equal to zero. Similarly, general function
minimization tools can be applied to try to minimize the sum of squares to
zero.

We therefore will seek to examine the solution of nonlinear equations problems
through tools for

\begin{itemize}
\item
  solution of nonlinear equations, generally through providing program code to
  evaluate the residuals \(r(x)\)
\item
  solution of nonlinear least squares problems by the same mechanism. However,
  in this case there may be more residuals than parameters, and the returned
  ``solution'' may not be have all residuals effectively zero
\item
  minimization of the sum of squares \(r'(x)*r(x)\). Again, unless the sum of
  squares is zero, we have not solved the nonlinear equations.
\end{itemize}

\section{Nonlinear equations tools}\label{nonlinear-equations-tools}

The software tools for nonlinear equations largely parallel those for nonlinear
function minimization and nonlinear least squares. Many of these have at least
a motivation in what is referred to as Newton's method (Polyak (\cite{Polyak2007}{2007}), Ypma (\cite{1995Ypma}{1995})).
This discussion
uses tools in R for approaching nonlinear equations problems. Unfortunately, R
is less well-provided with tools directly for nonlinear equations than, say,
Julia. (See, for example,
https://docs.sciml.ai/NonlinearSolve/stable/solvers/nonlinear\_system\_solvers/
Nevertheless, using the Consolidated R Archive Network repository ({https://cran.r-project.org/})
we do recognize the following R tools for nonlinear equations (there may be
others, and we welcome communications):

\begin{itemize}
\item
  package \texttt{nleqslv} uses Newton or Broyden methods to try to solve a set of
  nonlinear equations, of which Broyden is the default.
  Within each general approach, the stepsize is adjusted
  by several ``global'' strategies, of which the ``double dogleg'' strategy of
  J. E. Dennis and Schnabel (\cite{DenSchnab83}{1983}) is the default.
\item
  function \texttt{dfsane()} from package \texttt{BB} (Varadhan and Gilbert (\cite{p-BB}{2009})), which uses a projected gradient
  approach of LaCruz, Martinez, and Raydan (\cite{LaCruzEtal2006}{2006})
\item
  package \texttt{dfsaneacc} is an independent implementation of the same approach as
  \texttt{BB::dfsane()} (Birgin et al. (\cite{p-dfsaneacc}{2024}))
\end{itemize}

We note that the package \texttt{ktsolve} (\textbf{Configurable function for solving families of
nonlinear equations}) uses solvers from \texttt{BB} and \texttt{nleqslv}.

\section{Nonlinear least squares tools}\label{nonlinear-least-squares-tools}

R is quite well-served for tools for the related tasks of nonlinear least squares
(the minimization of the sum of squares) and nonlinear regression (where we seek
the model parameters and their properties). Base R (R Development Core Team (\cite{Rcite}{2008})) has the \texttt{nls()} function
which, while now quite dated, still has very wide-ranging capability, though it
lacks a stabilized (i.e., Levenberg-Marquardt) solver. By contrast, both packages
\texttt{minpack.lm} (Mullen, Elzhov, and Bolker (\cite{minpack.lm}{2012})) and \texttt{nlsr} (John C. Nash and Murdoch (\cite{nlsr17}{2017})) have stabilized solvers. Moreover,
\texttt{nlsr} attempts to use analytic or automatic differentiation to get derivatives
required by the solver. The forthcoming article John C. Nash and Bhattacharjee (\cite{NLSCompare23}{2023}) and the slightly
more general John C. Nash (\cite{Nash_2022Wires}{2022}) provide discussion and comparison of the tools.

\section{Nonlinear function minimization tools}\label{nonlinear-function-minimization-tools}

There are quite a lot of R tools for function minimization. This is quite often
called ``optimization'' within the R community, though strictly \textbf{function minimization}
is a sub-class of problems in optimization. For a very good overview, the
\emph{CRAN Task View: Optimization and Mathematical Programming} by Schwendinger and Borchers (\cite{CTVOptimization}{2024}) is
highly recommended.

For this article, we focus our attention to the tools available in package
\texttt{optimx} (John C. Nash and Varadhan (\cite{jnrv2011JSSOBKv43i09}{2011})). For solvers that seek local minima of functions
of several parameters, \texttt{optimx} offers more than enough choices. It is, however,
possible that we might want to explore problems that have multiple solutions,
that is, multiple minima of the sum of squares function, hopefully with zero
residuals. For such cases, it seems sensible to look at global or stochastic
minimization solvers (John C. Nash (\cite{nlpor14}{2014, chap. 15}), Cortez (\cite{2014Cortez}{2014})). We explore such ideas
briefly, but note that a wrapper for such methods to allow convenient use via
a unified call as in package \texttt{optimx} is not yet openly available.

\section{Dennis, Gay, Vu test problem}\label{dennis-gay-vu-test-problem}

This problem is described, with Fortran code, in John E. Dennis, Gay, and Vu (\cite{DenGayVu1983}{1983}). The problem,
in its original statement, has 8 parameters, but two of the equations (i.e.,
residual functions) are linear in the parameters, and a \textbf{reduced} problem
with 6 parameters and 6 residual functions can be easily set up. The technical
report provides five sets of data, that is, five sample problems. Also the
report suggests starting sets of parameters, but also looks at scaling these
``starts'' by 10 and 100. It seems that this problem is relatively difficult to
solve.

We have coded the full and reduced problems in R as \texttt{dgvf.R} and \texttt{dgvr.R}, with
a function \texttt{dgbprep()} that takes as its argument the ``Problem ID'' for each of
the suggested data sets. The technical report also suggests solutions. The
problem ID values are ``791129'', ``791226'', ``0121a'', ``0121b'' and ``0121c''.
We also wrote \texttt{dgvreduce()} and \texttt{dgvunreduce()} to
convert parameter vectors from the full to reduced problem or vice-versa.
All codes are provided in the Appendices before the References.

While we will not include the output here, we can confirm that running
the program in file \texttt{dgvtry.R} (via the
\texttt{source()} command) gives residuals that
are essentially zero for the proposed solution vectors
\texttt{xstar} or \texttt{xstrr} for all five problems in both full and reduced forms.

Now let us attempt some solutions with R tools for the example problem ``0121a''.

\subsection{DGV with nleqslv}\label{dgv-with-nleqslv}

For the default method and global strategy, neither full nor reduced solutions are
acceptable from starting parameters (\texttt{x0} for the full problem). However, we see
that the termination code (\texttt{termcd}) is not 1, so the software is informing us that
a satisfactory solution has not been found.

\tiny
\begin{verbatim}
#> Prepare Dennis-Gay-Vu problem for id= 0121a
\end{verbatim}

\begin{verbatim}
#> Using dgvf()
\end{verbatim}

\begin{verbatim}
#> Solution:[1]   0.002468503  -0.825902531   0.002126131  -0.020796338  -2.648814611
#> [6]   2.241598979 -19.644981645   0.708379863
#> SS= 0.005900999  termination code= 4
\end{verbatim}

\begin{verbatim}
#> NULL
\end{verbatim}

\begin{verbatim}
#> Using dgvr()
\end{verbatim}

\begin{verbatim}
#> Solution:[1]   0.002810679   0.001124791  -0.444240100   2.245941731 -20.424571868
#> [6]   0.744305155
#> SS= 0.006554426  termination code= 4
\end{verbatim}

\begin{verbatim}
#> NULL
\end{verbatim}
\normalsize

We can try different options for method and step length, and package \texttt{nleqslv} has a
tool, \texttt{testnslv()}, to allow this to be carried out easily.

\tiny
\begin{verbatim}
#> Prepare Dennis-Gay-Vu problem for id= 0121a
\end{verbatim}

\begin{verbatim}
#>  Using dgvf()
\end{verbatim}

\begin{verbatim}
#> Call:
#> testnslv(x = x0, fn = dgvf, rhs = sigma)
#> 
#> Results:
#>     Method Global termcd Fcnt Jcnt Iter Message     Fnorm
#> 1   Newton  cline      1   72   38   38   Fcrit 1.153e-27
#> 2   Newton  qline      1   72   38   38   Fcrit 1.153e-27
#> 3   Newton  gline      1   75   24   24   Fcrit 2.888e-22
#> 4   Newton pwldog      1  192  136  136   Fcrit 1.459e-29
#> 5   Newton dbldog      1   89   83   83   Fcrit 1.599e-28
#> 6   Newton   hook      1  140  132  132   Fcrit 6.456e-18
#> 7   Newton   none      1   19   19   19   Fcrit 2.593e-28
#> 8  Broyden  cline      1  102    4   46   Fcrit 2.669e-19
#> 9  Broyden  qline      1  109    5   50   Fcrit 5.047e-22
#> 10 Broyden  gline      1  130    4   38   Fcrit 1.968e-19
#> 11 Broyden pwldog      4  214    8  150 Maxiter 5.835e-03
#> 12 Broyden dbldog      4  214    7  150 Maxiter 2.950e-03
#> 13 Broyden   hook      4  241    7  150 Maxiter 1.445e-02
#> 14 Broyden   none      4   20    1   20 Maxiter 1.673e+04
\end{verbatim}

\begin{verbatim}
#> 
#>  Using dgvr()
\end{verbatim}

\begin{verbatim}
#> Call:
#> testnslv(x = x0, fn = dgvf, rhs = sigma)
#> 
#> Results:
#>     Method Global termcd Fcnt Jcnt Iter Message     Fnorm
#> 1   Newton  cline      1   72   38   38   Fcrit 1.153e-27
#> 2   Newton  qline      1   72   38   38   Fcrit 1.153e-27
#> 3   Newton  gline      1   75   24   24   Fcrit 2.888e-22
#> 4   Newton pwldog      1  192  136  136   Fcrit 1.459e-29
#> 5   Newton dbldog      1   89   83   83   Fcrit 1.599e-28
#> 6   Newton   hook      1  140  132  132   Fcrit 6.456e-18
#> 7   Newton   none      1   19   19   19   Fcrit 2.593e-28
#> 8  Broyden  cline      1  102    4   46   Fcrit 2.669e-19
#> 9  Broyden  qline      1  109    5   50   Fcrit 5.047e-22
#> 10 Broyden  gline      1  130    4   38   Fcrit 1.968e-19
#> 11 Broyden pwldog      4  214    8  150 Maxiter 5.835e-03
#> 12 Broyden dbldog      4  214    7  150 Maxiter 2.950e-03
#> 13 Broyden   hook      4  241    7  150 Maxiter 1.445e-02
#> 14 Broyden   none      4   20    1   20 Maxiter 1.673e+04
\end{verbatim}
\normalsize

Running the multiple solutions as \texttt{testnslv()} does is possibly inefficient. In
preparing this vignette, the
following crude script was devised to exit after the first satisfactory solution has
been achieved. The code to do this follows.

\tiny
\begin{verbatim}
seqnslv <- function(x, fn, jac=NULL, ...,
                method=c("Newton", "Broyden"),
                global=c("qline", "cline", "gline", "pwldog", "dbldog", "hook", "none"))
{
    methods <- method
    globals <- global
    reslist<-list()
    doneOK <- FALSE # JN: used to break out of loops
    for(m in methods) {
        cat("method ",m,"\n")
        for(g in globals) {
            cat("global ",g,"\n")
            res <- try(nleqslv(x, fn, jac, ..., method=m, global=g),silent=TRUE)
            if(inherits(res,"try-error")) {
                cat("method ",m," global ",g," FAILED\n")
            } else {
              ss <- sum((res$fvec)^2)
              cat("method ",m," global ",g," termcd=",res$termcd," sumsq=",ss,"\n")
              if (res$termcd ==1){
                doneOK<- TRUE
                attr(res,"method") <- m
                attr(res,"global") <- g
                break
              }
            }
        } # end g
        if (doneOK) break
    } # end m
    res
}
\end{verbatim}
\normalsize

This approach gives the following result

\tiny
\begin{verbatim}
#> Prepare Dennis-Gay-Vu problem for id= 0121a
\end{verbatim}

\begin{verbatim}
#> Using dgvf()
\end{verbatim}

\begin{verbatim}
#> method  Newton 
#> global  qline 
#> method  Newton  global  qline  termcd= 1  sumsq= 2.305261e-27
\end{verbatim}

\begin{verbatim}
#> Solution:[1]  3.099869e-03 -8.190999e-01 -2.239405e-04 -1.677606e-02  2.681514e+00
#> [6]  2.250216e+00 -2.024170e+01  7.970983e-01
#> SS= 2.305261e-27  termination code= 1
\end{verbatim}

\begin{verbatim}
#> NULL
\end{verbatim}

\begin{verbatim}
#> Using dgvr()
\end{verbatim}

\begin{verbatim}
#> method  Newton 
#> global  qline 
#> method  Newton  global  qline  termcd= 1  sumsq= 4.576478e-27
\end{verbatim}

\begin{verbatim}
#> Solution:[1]  3.099869e-03 -2.239405e-04  2.681514e+00  2.250216e+00 -2.024170e+01
#> [6]  7.970983e-01
#> SS= 4.576478e-27  termination code= 1
\end{verbatim}

\begin{verbatim}
#> NULL
\end{verbatim}
\normalsize

\subsection{DGV with BB::dfsane and dfsaneacc}\label{dgv-with-bbdfsane-and-dfsaneacc}

The spectral gradient routines in package \texttt{BB::dfsane()} 
and in package \texttt{dfsaneacc} are intended to solve nonlinear
equations problems, but unfortunately do not seem
to give satisfactory solutions with the default options. Even with some
variation of the options available to each program, it seems difficult to
get a solution. \texttt{dfsaneacc} was able to get to a solution after over 100,000
iterations and a quite large setting (25) in the number of previous iterates
considered in the secant acceleration update (default is 6, which was much
slower).

\tiny
\begin{verbatim}
#> Prepare Dennis-Gay-Vu problem for id= 0121a
#> Warning in dfsane(x0, dgvf, quiet = TRUE, rhs = sigma): Unsuccessful
#> convergence.
#> dsol$par  for fn= dgvf   SS= 274.2222  at 
#> [1] -0.42517901 -0.77500000  0.06898961 -0.03542919 -2.65623588  2.50630841
#> [7] -0.31907139 -0.13885158
#> residuals:[1]  -0.38417901   0.05056042   1.03007993   0.72522327  -3.11248693
#> [6]   3.33557503   9.92863312 -12.37310089
#> Warning in dfsane(x0r, dgvr, quiet = TRUE, rhs = sigma): Unsuccessful
#> convergence.
#> dsolr$par  for fn= dgvr   SS= 151.8927  at 
#> [1] -0.56678206 -0.02359818 -2.60639467  2.65553824 -1.91014963  1.83848441
#> residuals:[1]  2.584243  1.457483  2.312137 -4.867993  8.962342 -5.807173
#> Prepare Dennis-Gay-Vu problem for id= 0121a
#> daasolx  for fn= dgvf   SS= 1.324049e-12  at 
#> [1]  3.099812e-03 -8.190991e-01 -2.239393e-04 -1.677666e-02  2.681530e+00
#> [6]  2.250219e+00 -2.024183e+01  7.970978e-01
#> residuals:[1]  6.647103e-07 -6.004497e-07 -6.437725e-07  2.965854e-07  1.354081e-07
#> [6] -2.972639e-08 -3.730332e-10  6.652416e-09
#> 2e+05  iterations
#> daasolrx  for fn= dgvr   SS= 5.900543e-13  at 
#> [1]  3.099867e-03 -2.239404e-04  2.681515e+00  2.250216e+00 -2.024171e+01
#> [6]  7.970983e-01
#> residuals:[1] -6.130108e-08  3.162569e-08 -1.747531e-08 -4.538748e-09 -1.037709e-07
#> [6]  7.577611e-07
#> 118611  iterations
5\end{verbatim}
\normalsize

\subsection{DGV with nlsr::nlfb()}\label{dgv-with-nlsrnlfb}

\tiny
\begin{verbatim}
#> Prepare Dennis-Gay-Vu problem for id= 0121a
#> Using dgvf()
#> nlfbsol <- nlfb(x0, dgvf, jac="jacentral", rhs=sigma)
#> residual sumsquares =  3.2693e-17  on  8 observations
#>     after  190    Jacobian and  263 function evaluations
#>   name            coeff          SE       tstat      pval      gradient    JSingval   
#> p1            0.00309987           Inf          0        NaN  -3.093e-06        8523  
#> p2               -0.8191           Inf          0        NaN  -5.002e-09        8523  
#> p3          -0.000223941           Inf          0        NaN   1.678e-06       7.622  
#> p4            -0.0167761           Inf          0        NaN  -2.507e-09       7.622  
#> p5               2.68151           Inf          0        NaN   4.708e-10      0.8517  
#> p6               2.25022           Inf          0        NaN   4.102e-09      0.8517  
#> p7              -20.2417           Inf          0        NaN   1.588e-09    0.009023  
#> p8              0.797098           Inf          0        NaN   4.051e-09    0.009023

#> Using dgvr()
#> nlfbsolr <- nlfb(x0r, dgvr, jac="jacentral", rhs=sigma)
#> residual sumsquares =  6.05e-16  on  6 observations
#>     after  209    Jacobian and  290 function evaluations
#>   name            coeff          SE       tstat      pval      gradient    JSingval   
#> p1            0.00309987           Inf          0        NaN   1.934e-05        8515  
#> p2           -0.00022394           Inf          0        NaN  -6.953e-06        8515  
#> p3               2.68151           Inf          0        NaN  -1.651e-09       4.198  
#> p4               2.25022           Inf          0        NaN   -2.01e-08       4.198  
#> p5              -20.2417           Inf          0        NaN  -9.281e-09     0.01397  
#> p6              0.797098           Inf          0        NaN  -2.706e-08     0.01397
\end{verbatim}
\normalsize

For both full and reduced problems we have good solutions. However, the
singular values of the Jacobian at the solution are extreme, underlining
the difficulty of solution of these problems. We have not explored the
use of \texttt{nls()} from base R nor the package \texttt{minpack.lm}.

\subsection{DGV with optimx function minimizers}\label{dgv-with-optimx-function-minimizers}

We will minimize the functions

\tiny
\begin{verbatim}
ssdgvf<-function(x, rhs){
   res<-dgvf(x,rhs)
   ss<-sum(res^2)
   ss
}
ssdgvr<-function(x, rhs){
   res<-dgvr(x,rhs)
   ss<-sum(res^2)
   ss
}
\end{verbatim}
\normalsize

Trying ``MOST'' methods in package \texttt{optimx} showed only method ``Rvmmin'' succeeded for
the full problem, while a half dozen gave solutions for the reduced problem. The
calls are

\tiny
\begin{verbatim}
library(optimx)
osol <- opm(x0, ssdgvf, gr="grcentral", method="MOST",
            control=list(maxit=20000, maxfeval=25000), rhs=sigma)
osolr <- opm(x0r, ssdgvr, gr="grcentral", method="MOST", 
             control=list(maxit=20000, maxfeval=25000),rhs=sigma)
\end{verbatim}
\normalsize

To keep the display tidy, we show the results for seven methods, namely

\tiny
\begin{verbatim}
cmeth <- c("Rvmmin",  "nlm",  "ucminf",  "slsqp",  "ncg",  "Rcgmin, "mla")
\end{verbatim}
\normalsize

and limit the output to the two first parameters.

\tiny
\begin{verbatim}
#> Prepare Dennis-Gay-Vu problem for id= 0121a
#> Using dgvf()
#>                 p1 s1       p2 s2      value fevals gevals hevals conv  kkt1
#> Rvmmin  3.0999e-03    -0.81910    7.1117e-17   5616   4270      0    0  TRUE
#> Rcgmin  2.5794e-03    -0.82092    3.8466e-04  60003  17967      0    1 FALSE
#> mla     2.5202e-03    -0.82355    6.3165e-04   2543   6800      0    0 FALSE
#> ncg     1.7646e-03    -0.82320    2.2800e-03  60003  16682      0    1 FALSE
#> ucminf  3.1709e-06    -0.72498    2.0638e-02  20000  20000      0    3 FALSE
#> slsqp  -1.6049e-05    -0.72611    2.3149e-02   3453   3452      0    0 FALSE
#> nlm    -5.3352e-06    -0.71878    2.5649e-02  56585  56585      0    1 FALSE
#>         kkt2  xtime
#> Rvmmin FALSE  1.044
#> Rcgmin FALSE  4.257
#> mla    FALSE  1.374
#> ncg    FALSE  4.010
#> ucminf FALSE  4.008
#> slsqp  FALSE  0.717
#> nlm    FALSE 11.300
#> Using dgvr()
#>                 p1 s1          p2 s2      value fevals gevals hevals conv  kkt1
#> Rvmmin  3.0999e-03    -2.2394e-04    3.0171e-17   2706   2071      0    0  TRUE
#> Rcgmin  3.0999e-03    -2.2394e-04    8.1709e-15   7002   2201      0    0  TRUE
#> ncg     3.0998e-03    -2.2380e-04    1.9253e-11   7554   2220      0    0  TRUE
#> mla     3.2226e-03     2.6311e-04    2.4318e-04   1816   3016      0    0 FALSE
#> nlm    -2.6134e-05     8.3822e-06    6.1437e-02   1739   1739      0    0 FALSE
#> slsqp   1.1367e-12     1.1110e-11    7.1366e-02   8338   8337      0    0 FALSE
#> ucminf  7.1861e-08    -2.2112e-07    7.2835e-02  20000  20000      0    3 FALSE
#>         kkt2 xtime
#> Rvmmin FALSE 0.561
#> Rcgmin FALSE 0.595
#> ncg    FALSE 0.595
#> mla    FALSE 0.683
#> nlm    FALSE 0.405
#> slsqp  FALSE 1.910
#> ucminf FALSE 4.385
\end{verbatim}
\normalsize

Note that parameter scaling seems to give slightly better results.
However, we have not explored this possibility further.

\tiny
\begin{verbatim}
osolrs <- opm(x0r, ssdgvr, gr="grcentral", method=cmeth, 
              control=list(maxit=20000, maxfeval=25000, 
       parscale=c(.001, .0001, 1, 1, 10, 1)),rhs=sigma)
cat("Using dgvr()\n")
#> Using dgvr()
summary(osolrs, order=value, par.select=1:2)
#>               p1 s1          p2 s2      value fevals gevals hevals conv  kkt1
#> Rvmmin 0.0030999    -2.2394e-04    1.4274e-17   2772   2101      0    0  TRUE
#> nlm    0.0030999    -2.2394e-04    2.1059e-17   3136   3136      0    0  TRUE
#> ucminf 0.0030999    -2.2394e-04    4.5830e-17  19250  19250      0    0  TRUE
#> slsqp  0.0030999    -2.2394e-04    9.4500e-17  21395  21394      0    0  TRUE
#> ncg    0.0030999    -2.2102e-04    8.3162e-09  14872   5054      0    0 FALSE
#> Rcgmin 0.0031000    -2.2021e-04    1.3579e-08  13030   4621      0    0 FALSE
#> mla    0.0031172    -2.8545e-05    3.7627e-05   4816   4732      0    0 FALSE
#>         kkt2 xtime
#> Rvmmin FALSE 0.573
#> nlm    FALSE 0.710
#> ucminf FALSE 4.276
#> slsqp  FALSE 4.789
#> ncg    FALSE 1.310
#> Rcgmin FALSE 1.191
#> mla    FALSE 1.106
\end{verbatim}
\normalsize

\subsection{Residual scaling}\label{residual-scaling}

If we look at the residuals from the starting value \(x0\), the magnitudes
of the elements vary quite widely. Multiplying the residuals by a scaling
to equalize the contribution from each might help, but the following
experiments were unsuccessful. We use the following calls on the reduced
problem only and will not include the output here.

\tiny
\begin{verbatim}
rscale<-c(1e8, 1e8, 1e3, 1e3, 1, 1)
ssdgvrrs<-function(x, rhs){
   res<-dgvr(x,rhs)*rscale
   ss<-sum(res^2)
   ss
}
cat("Using dgvr() with residual scaling\n")
osolrrs <- opm(x0r, ssdgvrrs, gr="grcentral", method=cmeth, 
               control=list(maxit=20000, maxfeval=25000),rhs=sigma)
summary(osolrrs, order=value)
nmeth<-length(cmeth)
npar<-length(x0r)
for (ii in 1:nmeth){
  par<-as.numeric(osolrrs[ii, 1:npar])
  cat(cmeth[ii],"   ssdgvr = ",ssdgvr(par, rhs=sigma),"\n")
}
cat("Using dgvr() with residual and parameter scaling\n")
osolrpsrs <- opm(x0r, ssdgvrrs, gr="grcentral", method=cmeth, 
                 control=list(maxit=20000, maxfeval=25000, 
                              parscale=c(.001, .0001, 1, 1, 10, 1)),rhs=sigma)
summary(osolrpsrs, order=value)
for (ii in 1:nmeth){
  par<-as.numeric(osolrrs[ii, 1:npar])
  cat(cmeth[ii],"   ssdgvr = ",ssdgvr(par, rhs=sigma),"\n")
}
\end{verbatim}
\normalsize

\subsection{DGV with stochastic function minimizers}\label{dgv-with-stochastic-function-minimizers}

In 2017 Hans Werner Borchers prepared a prototype wrapper for stochastic
and global optimization solvers under the name \texttt{gloptim}. A Google Summer of
Code project in 2022 was started under the name \texttt{stochoptim} to add better
checks and documentation to these ideas, but this effort was abandoned
due to difficulties in effectively communicating with the student. One of
us has resurrected the \texttt{gloptim} package. (Several of the solvers are no
longer in the active CRAN repository and needed to be de-archived.)

We tried a small selection of such minimizers with rather unsatisfactory
results. We will include the code to run the examples and a saved version
of the output, as running times are quite long.

\tiny
\begin{verbatim}
# dgvglopt.R
library(gloptim)
source("dgvprep.R")
source("dgvf.R")
source("nleqsum.R")
source("chknleqsol.R")
pid <- "0121a"
prob<-dgvprep(pid)
sigma<-prob$sigma
x0<-prob$x0
source("dgvreduce.R")
source("dgvr.R")
x0r<-dgvreduce(x0)
shift <- 100
lbf <- x0 - shift
ubf <- x0 + shift
lbr <- x0r - shift
ubr <- x0r +shift 
dgvfss<-function(x){
  res<-dgvf(x, rhs=sigma)
  val<-sum(res^2)
}
dgvrss<-function(x){
  res<-dgvr(x, rhs=sigma)
  val<-sum(res^2)
}
  
tgf <- glopt(dgvfss, lb=lbf, ub=ubf, incl=c("deoptimr", "genoud", "ga", "pso"))
print(tgf)
tgr <- glopt(dgvrss, lb=lbr, ub=ubr, incl=c("deoptimr", "genoud", "ga", "pso"))
print(tgr)
\end{verbatim}

The results and timings are, unfortunately in this case, uninspiring.

\begin{verbatim}
> # dgvglopt.R
> library(gloptim)
> source("dgvprep.R")
> source("dgvf.R")
> source("nleqsum.R")
> source("chknleqsol.R")
> pid <- "0121a"
> prob<-dgvprep(pid)
Prepare Dennis-Gay-Vu problem for id= 0121a 
> sigma<-prob$sigma
> x0<-prob$x0
> source("dgvreduce.R")
> source("dgvr.R")
> x0r<-dgvreduce(x0)
> shift <- 100
> lbf <- x0 - shift
> ubf <- x0 + shift
> lbr <- x0r - shift
> ubr <- x0r +shift 
> dgvfss<-function(x){
+   res<-dgvf(x, rhs=sigma)
+   val<-sum(res^2)
+ }
> dgvrss<-function(x){
+   res<-dgvr(x, rhs=sigma)
+   val<-sum(res^2)
+ }
> tgf <- glopt(dgvfss, lb=lbf, ub=ubf, incl=c("deoptimr", "genoud", "ga", "pso"))
glopt: control=NULL
> print(tgf)
    method          fmin   time         p1           p2           p3         p4         p5          p6         p7
1       ga -5.916629e+17  5.193 -0.8917380  -0.06962564  -2.33956716   1.412347  0.7197578  -0.2825009 -1.4710387
2      pso  3.811090e-01  0.640  2.3000552  -4.95105203  -6.48411329   3.948841  2.0430412   2.3940513  0.8257454
3 deoptimr  2.751122e+00  5.143 99.4039055 -99.90418468 -98.99600227 -98.101759 96.4349572 102.0188937 98.6155200
4   genoud  2.477102e+01 29.179  0.0938765  -1.12740949   0.01184009   0.275271  0.9978030   1.2503798 -6.1394841
           p8
1  -2.4391020
2   0.3536293
3 -98.7978192
4   0.6878692

> tgr <- glopt(dgvrss, lb=lbr, ub=ubr, incl=c("deoptimr", "genoud", "ga", "pso"))
glopt: control=NULL
> print(tgr)
    method          fmin  time           p1           p2           p3          p4         p5          p6
1       ga -6.754425e+17 3.517 -0.009364945   0.04527876   -3.3798955   2.0277091  -6.843039   0.3015986
2 deoptimr  3.992541e-01 3.089 -1.677136320  -0.07111157    1.1226002   1.3877193  -6.418279  -8.1433227
3      pso  3.158459e+00 0.494 99.166368787 -99.33306963 -101.9835323 101.8854603 -99.791963 100.6789897
4   genoud  3.516621e+02 3.131  0.456135104   0.35255385    0.1855194   0.9448873  -3.225310  -0.7321973
\end{verbatim}

\section{nleqslv test problem}\label{nleqslv-test-problem}

Package \texttt{nleqslv} includes a test problem which we use here for illustration.

\[    r_1(x) = x_1^2 + x_2^2 -2 = 0  \]
\[    r_2(x) = exp(x_1 - 1) + x_2^3 - 2 = 0 \]

This seems to be a problem that is relatively easy to solve by most methods,
though we note that some approaches perform poorly. Notably,
a number of function minimization solvers find a local minimum that is not
a solution to the nonlinear equations. This is indicated by the residual
sum of squares value near 0.18336, with parameters near the suggested start

\texttt{xstart3\textless{}-c(1.48508,\ -1.0886e-06)}

We have suppressed most of the output for space reasons.

\begin{verbatim}
#> Sumsquares results of methods
#>            method   start                   ss
#> 1         nleqslv xstart1 7.91546228567908e-20
#> 2         nleqslv xstart2 6.47806667252515e-18
#> 3         nleqslv xstart3    0.183361725333783
#> 4      BB::dfsane xstart1    2.40003077549e-17
#> 5      BB::dfsane xstart2 2.44274187984104e-19
#> 6      BB::dfsane xstart3 1.25066039269404e-14
#> 7       dfsaneacc xstart1 1.67101720053509e-21
#> 8       dfsaneacc xstart2 1.50592337150088e-16
#> 9       dfsaneacc xstart3 1.00022577995405e-19
#> 10 nlfb-jacentral xstart1 1.35165621000982e-15
#> 11 nlfb-jacentral xstart2 9.76473803243686e-18
#> 12    Nelder-Mead xstart1 1.66815320909569e-08
#> 13            nlm xstart1 5.65111729255102e-16
#> 14         nlminb xstart1    0.183361654679628
#> 15       lbfgsb3c xstart1 1.27770986049496e-14
#> 16         Rtnmin xstart1 5.11399868633245e-15
#> 17            spg xstart1 2.03208269671654e-17
#> 18         ucminf xstart1 1.42472672550097e-19
#> 19         bobyqa xstart1 3.02237589087772e-14
#> 20           nmkb xstart1    0.183361747085402
#> 21        subplex xstart1    0.183361654677934
#> 22            ncg xstart1 6.54745022259549e-15
#> 23         Rcgmin xstart1 8.47277001984322e-21
#> 24            nvm xstart1 1.18526351009457e-24
#> 25         Rvmmin xstart1 1.18526351009457e-24
#> 26            mla xstart1 1.51249045387053e-10
#> 27          slsqp xstart1  1.2593529121072e-17
#> 28          tnewt xstart1 1.66994916589703e-24
#> 29       pracmanm xstart1    0.183361654677934
#> 30           nlnm xstart1                    0
#> 31    Nelder-Mead xstart2 9.14907810780667e-08
#> 32            nlm xstart2    0.183361654679974
#> 33         nlminb xstart2    0.183361654678151
#> 34       lbfgsb3c xstart2    0.183361654748689
#> 35         Rtnmin xstart2 1.71720848945297e-18
#> 36            spg xstart2    0.183361654677936
#> 37         ucminf xstart2    0.183361654677995
#> 38         bobyqa xstart2    0.183361654678425
#> 39           nmkb xstart2    0.183362331170298
#> 40        subplex xstart2    0.183361654677934
#> 41            ncg xstart2     0.18336165470041
#> 42         Rcgmin xstart2    0.183361654738436
#> 43            nvm xstart2    0.183361654678022
#> 44         Rvmmin xstart2    0.183361654678022
#> 45            mla xstart2 6.69885437003902e-09
#> 46          slsqp xstart2    0.183361654686446
#> 47          tnewt xstart2    0.183361654732427
#> 48       pracmanm xstart2    0.183361654677934
#> 49           nlnm xstart2 5.40241409895111e-13
#> 50    Nelder-Mead xstart3    0.183361654694565
#> 51            nlm xstart3    0.183361654679259
#> 52         nlminb xstart3    0.183361654680113
#> 53       lbfgsb3c xstart3 1.58628930536124e-15
#> 54         Rtnmin xstart3    0.183361654678318
#> 55         ucminf xstart3    0.183361654694565
#> 56         bobyqa xstart3    0.183361654677934
#> 57           nmkb xstart3  1.5183070349323e-06
#> 58        subplex xstart3 1.97215226305253e-31
#> 59            ncg xstart3    0.183361654681022
#> 60         Rcgmin xstart3      0.1833616546784
#> 61            nvm xstart3    0.183361654679689
#> 62         Rvmmin xstart3    0.183361654679689
#> 63          slsqp xstart3    0.183361654678086
#> 64          tnewt xstart3    0.183361654681015
#> 65       pracmanm xstart3 3.51800409292362e-26
#> 66           nlnm xstart3    0.183361654678596
#> 67 nlfb-jacentral xstart3    0.183361654678426
#> 68     nlfb-jafwd xstart2    0.183361654678427
#> 69    nlfb-jaback xstart3    0.183361654678425
#> 70      nlfb-jand xstart3    0.183361654677938
#> 71            ncg xstart3    0.183361654681022
#> 72            nvm xstart3    0.183361654679689
#> 73       lbfgsb3c xstart3 1.58628930536124e-15
#> times for different methods (10 repetitions)
#>                method   start    tmin     tmean    tmax
#> 1             nleqslv xstart1  109936  120334.8  196663
#> 2             nleqslv xstart2  112321  123596.4  178748
#> 3             nleqslv xstart3   88166   98686.5  159058
#> 4  lbfgsb3c-grcentral xstart1  822240  857358.3  985864
#> 5  lbfgsb3c-grcentral xstart2 1612390 1649005.5 1802940
#> 6  lbfgsb3c-grcentral xstart3 1076268   1111966 1280132
#> 7              dfsane xstart1  285125  303686.4  440083
#> 8              dfsane xstart2  344259    355297  419262
#> 9              dfsane xstart3  387176  397593.1  460252
#> 10               nlfb xstart1 1555174 1616051.3 1727092
#> 11               nlfb xstart2 1713308   1809668 2315558
#> 12               nlfb xstart3  327300  346383.9  440649
#> 13          dfsaneacc xstart1  710574  740496.5  819892
#> 14          dfsaneacc xstart2  526905    568740  764327
#> 15          dfsaneacc xstart3 1539579 1628672.8 1922037
\end{verbatim}
\normalsize

We note for this problem that most methods fail from \texttt{xstart3}, as the Jacobian at this point is
essentially singular and one component of the gradient is also small. Interestingly, quite large
changes to this particular start are needed to escape the local minimum, though some derivative-free
methods succeed in finding a solution. This is likely due to these methods taking large
trial steps in the parameters in early stages of their execution.

\section{trigexp test problem}\label{trigexp-test-problem}

The quite well-known \textbf{trigexp} test problem (see La Cruz and Raydan (\cite{LaCruzRaydan2003}{2003}, Problem 12))
is relatively easy to solve.

\tiny
\begin{verbatim}
trigexp <- function(x) {
  # Test function No. 12 in the Appendix of LaCruz and Raydan (2003)
  n <- length(x)
  F <- rep(NA, n)
  F[1] <- 3*x[1]^2 + 2*x[2] - 5 + sin(x[1] - x[2]) * sin(x[1] + x[2])
  tn1 <- 2:(n-1)
  F[tn1] <- -x[tn1-1] * exp(x[tn1-1] - x[tn1]) + x[tn1] * ( 4 + 3*x[tn1]^2) +
    2 * x[tn1 + 1] + sin(x[tn1] - x[tn1 + 1]) * sin(x[tn1] + x[tn1 + 1]) - 8 
  F[n] <- -x[n-1] * exp(x[n-1] - x[n]) + 4*x[n] - 3
  F
}
#> dfsane solution. SS = 2.5518e-13
#> nlsr::nlfb solution. SS = 2.916e-23
#> nleqslv solution. SS = 5.9479e-21
#> dfsaneacc solution. SS = 2.4239e-13
\end{verbatim}
\normalsize

\section{The Abbott and Brent problem}\label{the-abbott-and-brent-problem}

This problem seems to have first appeared as Example 7 in Abbott and Brent (\cite{AbbottBrent1975}{1975}), then
again as Example 1 in Alefeld, Gienger, and Potra (\cite{Alefeld94}{1994}) and yet later in the quite large collection of
problems Luksan and Vlcek (\cite{Luksan2003}{2018}). It is a relatively straightforward problem as seen by our
computational results below. The code for the residuals is as follows.

\tiny
\begin{verbatim}
brent <- function(x) {
# Luksan (2018) #63, Alefeld(1994) Example 1, Abbot & Brent (1975) Ex. 7
  n <- length(x)
  tnm1 <- 2:(n-1)
  F <- rep(NA, n)
  F[1] <- 3 * x[1] * (x[2] - 2*x[1]) + (x[2]^2)/4 
  F[tnm1] <-  3 * x[tnm1] * (x[tnm1+1] - 2 * x[tnm1] + x[tnm1-1]) + 
    ((x[tnm1+1] - x[tnm1-1])^2) / 4   
  F[n] <- 3 * x[n] * (20 - 2 * x[n] + x[n-1]) +  ((20 - x[n-1])^2) / 4
  F
}
#> Function at start has SS= 6159.9
#> dfsane default SS= 1.7461e-13
#> dfsane M=200  SS= 4.4505e-13
#> nleqslv  SS= 2.7741e-17
#> Check with nleqslv fvec SS= 2.7741e-17
#> nlfb SS= 3.2236e-15
#> dfsaneacc SS= 9.9399e-13
\end{verbatim}
\normalsize

Optimization solvers using the sum of squares give the following sample of
results. The stochastic optimizers, of which only four were chosen more or
less at random, were particularly unsuccessful and time-consuming, and
are deemed not worth reporting here, though code to run them is included
but commented-out in the example. Moreover, the solvers are such that some
cannot easily be limited by a maximum number of iterations or function counts.

\tiny
\begin{verbatim}
#> Function at start has SS= 6159.9
\end{verbatim}

\begin{verbatim}
#> Brent-Abbot problem
\end{verbatim}

\begin{verbatim}
#> Warning in BB::spg(par = spar, fn = efn, gr = egr, lower = slower, upper =
#> supper, : Unsuccessful convergence.
\end{verbatim}

\begin{verbatim}
#>                   p1 s1         p2 s2      value fevals gevals hevals conv
#> nlm      -0.00037063     0.0050205    6.2346e-13    419    419      0    0
#> nvm      -0.00076012    -0.0011585    2.3907e-12   1460    528      0    0
#> lbfgsb3c -0.03631404    -0.0872774    6.8173e-02   3501   3501      0    1
#> ncg       0.21993861     0.4476898    1.0761e+00   7630   3500      0    1
#> spg       0.35303095     0.6118932    1.1604e+00   4631   3503      0    1
#> nlminb    0.55820648     0.8947510    5.4060e+00    160    151      0    1
#> pracmanm  0.48442607     0.7474630    8.3493e+00  35053      0      0    1
#>           kkt1  kkt2 xtime
#> nlm       TRUE FALSE 0.613
#> nvm       TRUE FALSE 0.894
#> lbfgsb3c FALSE FALSE 5.275
#> ncg      FALSE FALSE 5.296
#> spg      FALSE FALSE 5.350
#> nlminb   FALSE FALSE 0.301
#> pracmanm FALSE FALSE 1.156
\end{verbatim}
\normalsize

\section{Observations}\label{observations}

The illustrative examples above suggest the following summary advice for workers
wanting to solve nonlinear equations:

\begin{itemize}
\item
  The speed of the methods was of the same general order of magnitude, except for
  stochastic optimizers which could take very much more time and effort.
\item
  Nonlinear equation tools, when they work, are easy to apply. \texttt{nleqslv}, moreover,
  had the fastest running time in the limited trials we made. This is not unexpected,
  since it calls code in Fortran, and these codes are tailored to the nonlinear
  equations problem.
\item
  Function minimization methods had the highest failure rate. Since we square the
  residuals, we are likely losing information useful to the minimization process.
  This also applies to stochastic solvers. There is, unfortunately, inconsistency
  of success -- sometimes normally ``good'' solvers do poorly and less well-regarded
  methods do well. We note that success is obvious if a zero sum of squared
  residuals is obtained.
\item
  Nonlinear least squares via \texttt{nlsr::nlfb()} was generally reliable. We
  automatically have the size of the residuals and their sum of squares as a
  measure of success. Moreover, the singular
  values of the Jacobian provided by tools in package \texttt{nlsr} are helpful
  to understand when problems may be
  difficult to solve. Note that computing the singular values may slow down the
  performance of this particular tool, which is, moreover, coded entirely in R.
\end{itemize}

\section{Appendices}\label{appendices}

The appendices provide code for some of the examples.

\subsection{dgvprep()}\label{dgvprep}

\tiny
\begin{verbatim}
# code for different DGV problems  -- dgvprep.R
dgvprep <- function(pid="791129", reduced=FALSE){
   snames <- c("sigmax", "sigmay", "sigmaa", "sigmab", "sigmac",
                "sigmad", "sigmae", "sigmaf")
#
#
   dostop<-FALSE
   cat("Prepare Dennis-Gay-Vu problem for id=",pid,"\n")
   if (pid == "791129"){
     sigma<-c(0.485, -0.0019, -0.0581, 0.015, 0.105, 0.0406, 0.167, -0.399)
     x0 <- c(0.299, 0.186, -0.0273, 0.0254, -0.474, 0.474, -0.0892, 0.0892)
     xstar<-c(-6.321349025e-3, 4.913213490e-1, -1.998156408e-3, 9.815640840e-5,
        1.226569755e-1, -1.003153205e-1, -4.023517593, -2.071785527e-2)
#
   } else if (pid == "791226"){
     sigma<-c(-0.69, -0.044, -1.57, -1.31, -2.65, 2.0, -12.6, 9.48)
     x0 <- c(-0.3, -0.39, 0.3, -0.344, -1.2, 2.69, 1.59, -1.5)
     xstar<-c(-3.116266056e-1, -3.783733944e-1, 3.282442301e-1, -3.722442301e-1,
        -1.282227094, 2.494300312, 1.554865879, -1.384637843)
# 
   } else if (pid == "0121a") {
     sigma<-c(-0.816, -0.017, -1.826, -0.754, -4.839, -3.259, -14.023, 15.467 )
     x0 <- c(-0.41, -0.775, 0.03, -0.047, -2.565, 2.565, -0.754, 0.754)
     xstar<-c(3.099869097e-3, -8.190998691e-1, -2.239405352e-4, -1.677605946e-2,
          2.681514498, 2.250215931, -2.024170463e+1, 7.970982952e-1)
# 
   } else if (pid == "0121b"){
     sigma<-c(-0.809, -0.021, -2.04, -0.614, -6.903, -2.934, -26.328, 18.639)
     x0 <- c(-0.056, -0.753, 0.026, -0.047, -2.991, 2.991, -0.568, 0.568)
     xstar<-c(9.034542990e-3, -8.180345430e-1, -4.450738446e-4, -2.055492616e-2,
             2.773429036, 2.529477259, -1.480097186e+1, 5.220468844e-1)
   } else if (pid == "0121c"){
     sigma<-c(-0.807, -0.021, -2.379, -0.364, -10.541, -1.961, -51.551, 21.053)
     x0 <- c(-0.074, -0.733, 0.013, -0.034, -3.632, 3.632, -0.289, 0.289)
     xstar<-c(5.140417418e-2, -8.584041742e-1, 1.047333626e-3, -2.204733363e-2,
              2.861205288, 2.949155438, -8.304243489, -1.454992413e-1)
   } else {
       cat("Unrecognized problem id\n")
       dostop<-TRUE
   } 
   if(dostop) stop("unrecognized problem")
   names(sigma)<-snames
   if (reduced) {
     xr <- rep(NA, 6)
     xr[1]<-x0[1]
     xr[2]<-x0[3]
     xr[3]<-x0[5]
     xr[4]<-x0[6]
     xr[5]<-x0[7]
     xr[6]<-x0[8]
     x0<-xr
     xs <- rep(NA, 6)
     xs[1]<-xstar[1]
     xs[2]<-xstar[3]
     xs[3]<-xstar[5]
     xs[4]<-xstar[6]
     xs[5]<-xstar[7]
     xs[6]<-xstar[8]
   }
   dgv <- list(pid=pid, sigma=sigma, x0=x0, xstar=xstar, reduced=reduced)
   dgv
}
\end{verbatim}
\normalsize

\subsection{dgvf()}\label{dgvf}

\tiny
\begin{verbatim}
dgvf <- function(x, rhs=RHS){ # Dennis, Gay, Vu NLLS/NLEQ prob
# John C. Nash 2024-4020
# From TR83 article:
# John Dennis, David Gay, Phuong Vu, 
# A new nonlinear equations test problem, 
# Technical Report 83-16, 
# Mathematical Sciences Department, 
# Rice University, 1983
#
# x = vector of 8 elements for full DGV problem
# Full problem:
# a=x[1], b=x[2], c=x[3], d=x[4]
# t=x[5], u=x[6], v=x[7], w=x[8]
# rhs=c(sigmax, sigmay, sigmaa, sigmab, sigmac,
#           sigmad, sigmae, sigmaf) (transposed)
#
  sigmax <- rhs[1]
  sigmay <- rhs[2]
  sigmaa <- rhs[3]
  sigmab <- rhs[4]
  sigmac <- rhs[5]
  sigmad <- rhs[6]
  sigmae <- rhs[7]
  sigmaf <- rhs[8]
#
   t2mv2 <- x[5]^2 - x[7]^2
   u2mw2 <- x[6]^2 - x[8]^2
   t2m3v2 <- x[5]^2 - 3.0*x[7]^2
   v2m3t2 <- x[7]^2 - 3.0*x[5]^2
   u2m3w2 <- x[6]^2 - 3.0*x[8]^2
   w2m3u2 <- x[8]^2 - 3.0*x[6]^2
   ctv <- x[3] * x[5] * x[7]
   duw <- x[4] * x[6] * x[8]
   atv <- x[1] * x[5] * x[7]
   buw <- x[2] * x[6] * x[8]
   at  <- x[1] * x[5]
   bu  <- x[2] * x[6]
   cv  <- x[3] * x[7]
   dw  <- x[4] * x[8]
   ct  <- x[3] * x[5]
   av  <- x[1] * x[7]
   du  <- x[4] * x[6]
   bw  <- x[2] * x[8]
#
   fvec<-rep(NA,8)
   fvec[1] <- x[1] + x[2] - sigmax
   fvec[2] <- x[3] + x[4] - sigmay
   fvec[3] <- at + bu - cv - dw -sigmaa
   fvec[4] <- av + bw + ct + du - sigmab
   fvec[5] <- x[1]*t2mv2 - 2.0*ctv +x[2]*u2mw2 -
             2.0*duw -sigmac
   fvec[6] <- x[3]*t2mv2 + 2.0*atv +x[4]*u2mw2 +
             2.0*buw -sigmad
   fvec[7] <- at*t2m3v2 + cv*v2m3t2 + bu*u2m3w2 +
             dw*w2m3u2 - sigmae
   fvec[8] <- ct*t2m3v2 - av*v2m3t2 + du*u2m3w2 -
             bw*w2m3u2 - sigmaf
   fvec # return
}
\end{verbatim}
\normalsize

\subsection{dgvr()}\label{dgvr}

\tiny
\begin{verbatim}
dgvr <- function(x, rhs=RHS){ # Dennis, Gay, Vu NLLS/NLEQ prob
# John C. Nash 2024-4020
# From TR83 article:
# John Dennis, David Gay, Phuong Vu, 
# A new nonlinear equations test problem, 
# Technical Report 83-16, 
# Mathematical Sciences Department, 
# Rice University, 1983
#
# x = vector of 8 elements for full DGV problem
# REDUCED PROBLEM 
# rhs=c(sigmax, sigmay, sigmaa, sigmab, sigmac,
#           sigmad, sigmae, sigmaf) (transposed)
#
  sigmax <- rhs[1]
  sigmay <- rhs[2]
  sigmaa <- rhs[3]
  sigmab <- rhs[4]
  sigmac <- rhs[5]
  sigmad <- rhs[6]
  sigmae <- rhs[7]
  sigmaf <- rhs[8]
#
# Reduced problem
# a=x[1], c=x[2], t=x[3], u=x[4], v=x[5], w=x[6]
   t2mv2 <- x[3]^2 - x[5]^2
   u2mw2 <- x[4]^2 - x[6]^2
   t2m3v2 <- x[3]^2 - 3.0*x[5]^2
   v2m3t2 <- x[5]^2 - 3.0*x[3]^2
   u2m3w2 <- x[4]^2 - 3.0*x[6]^2
   w2m3u2 <- x[6]^2 - 3.0*x[4]^2
   b <- sigmax - x[1]
   d <- sigmay - x[2]
   ctv <- x[2] * x[3] * x[5]
   duw <- d * x[4] * x[6]
   atv <- x[1] * x[3] * x[5]
   buw <- b * x[4] * x[6]
   at  <- x[1] * x[3]
   bu  <- b * x[4]
   cv  <- x[2] * x[5]
   dw  <- d * x[6]
   ct  <- x[2] * x[3]
   av  <- x[1] * x[5]
   du  <- d * x[4]
   bw  <- b * x[6]
#
   fvec<-rep(NA,6)
   fvec[1] <- at + bu - cv - dw - sigmaa
   fvec[2] <- av + bw + ct + du - sigmab
   fvec[3] <- x[1]*t2mv2 - 2.0*ctv + b*u2mw2 -
             2.0*duw - sigmac
   fvec[4] <- x[2]*t2mv2 + 2.0*atv + d*u2mw2 +
             2.0*buw - sigmad
   fvec[5] <- at*t2m3v2 + cv*v2m3t2 + bu*u2m3w2 +
             dw*w2m3u2 - sigmae
   fvec[6] <- ct*t2m3v2 - av*v2m3t2 + du*u2m3w2 -
             bw*w2m3u2 - sigmaf
   fvec # return
}
\end{verbatim}
\normalsize

\subsection{dgvreduce()}\label{dgvreduce}

\tiny
\begin{verbatim}
# code for different DGV problems  -- dgvreduce.R
dgvreduce <- function(xin){
# sigma NOT needed
     xr <- rep(NA, 6)
     xr[1]<-xin[1]
     xr[2]<-xin[3]
     xr[3]<-xin[5]
     xr[4]<-xin[6]
     xr[5]<-xin[7]
     xr[6]<-xin[8]
   xr
}
\end{verbatim}
\normalsize

\subsection{dgvunreduce()}\label{dgvunreduce}

\tiny
\begin{verbatim}
dgvunreduce <- function(xred, sigma){ # change from reduced to full
   snames <- c("sigmax", "sigmay", "sigmaa", "sigmab", "sigmac",
                "sigmad", "sigmae", "sigmaf")
#
#
     x <- rep(NA, 8)
     x[1]<-xred[1]
     x[2]<-sigma[1] - xred[1]
     x[3]<-xred[2]
     x[4]<-sigma[2] - xred[2]
     x[5]<-xred[3]
     x[6]<-xred[4]
     x[7]<-xred[5]
     x[8]<-xred[6]
   x
}
\end{verbatim}
\normalsize

\subsection{dgvtry()}\label{dgvtry}

\tiny
\begin{verbatim}
# code for different DGV problems  -- dgvprep.R
source("dgvprep.R")
source("dgvf.R")
source("dgvr.R")
source("dgvreduce.R")
source("dgvunreduce.R")

pidvec<-c( "791129", "791226", "0121a", "0121b", "0121c")
    
for (pid in pidvec){
  prob<-dgvprep(pid)
  cat("pid:",pid,"\n")
  sigma<-prob$sigma
  x0<-prob$x0
  xstar<-prob$xstar
  cat("rhs:"); print(sigma)
  cat("x0:");print(x0)
  val0 <- dgvf(x0, rhs=sigma)
  cat("dgvf(x0):"); print(val0)
  val10 <- dgvf(10*x0, rhs=sigma)
  cat("dgvf(10*x0):"); print(val10)
  val100 <- dgvf(100*x0, rhs=sigma)
  cat("dgvf(100*x0):"); print(val100)
  vstar<- dgvf(xstar, rhs=sigma)
  cat("dgvf(xstar):"); print(vstar)
  cat("SS=", sum(vstar^2),"\n")
  cat("\n\nREDUCED:\n")
  prob<-dgvprep(pid, reduced=TRUE)
  cat("pid:",pid,"\n")
  sigma<-prob$sigma
  x0r<-prob$x0
  xstrr<-dgvreduce(prob$xstar)
  cat("xstrr:"); print(xstar)
  cat("check unred:"); dgvunreduce(xstrr, sigma)
  cat("x0r:");print(x0r)
  val0r <- dgvr(x0r, rhs=sigma)
  cat("dgvr(x0r):"); print(val0r)
  val10r <- dgvr(10*x0r, rhs=sigma)
  cat("dgvr(10*x0r):"); print(val10r)
  val100r <- dgvr(100*x0r, rhs=sigma)
  cat("dgvr(100*x0r):"); print(val100r)
  vstrr<- dgvr(xstrr, rhs=sigma)
  cat("dgvr(xstrr):"); print(vstrr)
  cat("SS=", sum(vstrr^2),"\n")
  cat("unreduced xstrr:")
  xsu<-dgvunreduce(xstrr, sigma)
  print(xsu)
  fres<-dgvf(xsu, sigma)
  cat("full resids ="); print(fres)
  cat("SS=",sum(fres^2),"\n")
  cat("=====================================\n\n\n")
  
}
\end{verbatim}
\normalsize

\bibliographystyle{siam}

\bibliography{nleq}       

\begin{thebibliography}{10}

\bibitem{AbbottBrent1975}
{\sc J.~P. Abbott and R.~P. Brent}, {\em Fast local convergence with single and
  multistep methods for nonlinear equations}, The Journal of the Australian
  Mathematical Society. Series B. Applied Mathematics, 19 (1975), pp.~173 --
  199.

\bibitem{Alefeld94}
{\sc G.~Alefeld, A.~Gienger, and F.~Potra}, {\em Efficient validation of
  solutions of nonlinear systems}, SIAM Journal on Numerical Analysis, 31
  (1994), pp.~252--260.

\bibitem{p-dfsaneacc}
{\sc E.~G. Birgin, J.~L. Gardenghi, D.~S. Marcondes, and J.~M. Martinez}, {\em
  {dfsaneacc: Accelerated Derivative-Free Method for Large-Scale Nonlinear
  Systems of Equations}}, 2024.
\newblock R package version 1.0.2.

\bibitem{2014Cortez}
{\sc P.~Cortez}, {\em Modern Optimization with R}, Springer, Sept. 2014.

\bibitem{DenGayVu1983}
{\sc J.~E. Dennis, D.~M. Gay, and P.~A. Vu}, {\em A new nonlinear equations
  test problem}, tech. rep., Mathematical Sciences Department, Rice
  Universithy, 1983.
\newblock TR83-16.

\bibitem{DenSchnab83}
{\sc J.~E. Dennis and R.~B. Schnabel}, {\em Numerical Methods for Unconstrained
  Optimization and Nonlinear Equations}, Prentice-Hall, Englewood Cliffs, NJ,
  1983.

\bibitem{LaCruzRaydan2003}
{\sc W.~La~Cruz and M.~Raydan}, {\em Nonmonotone spectral methods for
  large-scale nonlinear systems}, {Optimization Methods \& Software}, 18
  (2003), pp.~583--599.

\bibitem{LaCruzEtal2006}
{\sc W.~LaCruz, J.~Martinez, and M.~Raydan}, {\em {Spectral residual method
  without gradient information for solving large-scale nonlinear systems of
  equations}}, Mathematics of Computation, 75 (2006), pp.~1429 -- 1448.

\bibitem{Luksan2003}
{\sc L.~Luksan and J.~Vlcek}, {\em Test problems for unconstrained
  optimization}, tech. rep., {Institute of Computer Science, Academy of
  Sciences of the Czech Republic}, Sept 2018.
\newblock Technical Report No. 897.

\bibitem{minpack.lm}
{\sc K.~M. Mullen, T.~V. Elzhov, and B.~Bolker}, {\em {minpack.lm: R interface
  to the Levenberg-Marquardt nonlinear least-squares algorithm found in
  MINPACK, plus support for bounds}}, 2012.
\newblock R package version 1.1-6.

\bibitem{nlpor14}
{\sc J.~C. Nash}, {\em {Nonlinear Parameter Optimization Using R Tools}}, John
  Wiley \& Sons: Chichester, May 2014.

\bibitem{Nash_2022Wires}
{\sc J.~C. Nash}, {\em Function minimization and nonlinear least squares in r},
  {WIREs} Computational Statistics,  (2022).
\newblock {JNfile: JNWireSOpt.pdf}.

\bibitem{NLSCompare23}
{\sc J.~C. Nash and A.~Bhattacharjee}, {\em {A comparison of R tools for
  nonlinear least squares modeling}}, {R Journal},  (2023).
\newblock {Accepted for publication in the R Journal, in November 2023.}

\bibitem{nlsr17}
{\sc J.~C. Nash and D.~Murdoch}, {\em {nlsr: Functions for Nonlinear Least
  Squares Solutions}}, 2017.
\newblock R package version 2017.2.19. Package contains 3 vignettes, "nlsr
  Background, Development, Examples and Discussion", "nlsr Derivatives", and
  "Specifying Fixed Parameters". {https://cran.r-project.org/package=nlsr}.

\bibitem{jnrv2011JSSOBKv43i09}
{\sc J.~C. Nash and R.~Varadhan}, {\em Unifying optimization algorithms to aid
  software system users: optimx for {R}}, Journal of Statistical Software, 43
  (2011), pp.~1--14.

\bibitem{NelsonHodgkin81}
{\sc C.~V. Nelson and B.~C. Hodgkin}, {\em Determination of magnitudes,
  directions, and locations of two independent dipoles in a circular conducting
  region from boundary potential measurements}, IEEE Transactions on Biomedical
  Engineering, BME-28 (1981), pp.~817--823.

\bibitem{Polyak2007}
{\sc B.~Polyak}, {\em Newton’s method and its use in optimization}, European
  Journal of Operational Research, 181 (2007), pp.~1086--1096.

\bibitem{Rcite}
{\sc {R Development Core Team}}, {\em {R}: A Language and Environment for
  Statistical Computing}, R Foundation for Statistical Computing, Vienna,
  Austria, 2008.
\newblock {ISBN} 3-900051-07-0.

\bibitem{CTVOptimization}
{\sc F.~Schwendinger and H.~W. Borchers}, {\em {CRAN Task View: Optimization
  and Mathematical Programming}}, mar 2024.
\newblock Version 2024-03-17.

\bibitem{p-BB}
{\sc R.~Varadhan and P.~Gilbert}, {\em {BB}: An {R} package for solving a large
  system of nonlinear equations and for optimizing a high-dimensional nonlinear
  objective function}, Journal of Statistical Software, 32 (2009), pp.~1--26.
\newblock R package version 2019.10-1.

\bibitem{1995Ypma}
{\sc T.~J. Ypma}, {\em {Historical development of the Newton-Raphson method}},
  SIAM Review, 37 (1995), pp.~531--551.

\end{thebibliography}

\section*{Addresses}

\noindent
John C. Nash\\
University of Ottawa (retired)\\%
Telfer School of Management\\ Ottawa, Ontario\\
\textit{ORCiD: {https://orcid.org/0000-0002-2762-8039}}\\%
{mailto:profjcnash@gmail.com}

\vspace{5mm}
\noindent
Ravi Varadhan\\
Johns Hopkins University\\%
Johns Hopkins University Medical School\\ Baltimore, Maryland\\
{mailto:ravi.varadhan@jhu.edu}

\end{document}